\documentclass[aps,showpacs,twocolumn]{revtex4}
\usepackage{graphicx}
\def\phii{$\Phi(z)$ }

\def\pt{$p_T$ }

\begin{document}
\title{Fractional Energy Loss and Centrality Scaling}
 \author{Rudolph C. Hwa$^1$ and  C.\ B.\ Yang$^{2}$}
\affiliation{$^1$Institute of Theoretical Science and Department of Physics\\
University of Oregon, Eugene, OR 97403-5203, USA\\
$^2$Institute of Particle Physics, Hua-Zhong Normal University,
Wuhan 430079, P.\ R.\ China}
\begin{abstract}
The phenomenon of centrality scaling in the high-\pt spectra of $\pi^0$
produced in Au-Au collisions at $\sqrt s=200$ GeV is examined in
the framework of  relating fractional energy loss to fractional
centrality increase. A new scaling behavior is found where the scaling
variable is given a power-law dependence on $N_{\rm part}$. The exponent
$\gamma$ specifies the fractional proportionality relationship between
energy loss and centrality, and is a phenomenologically determined
number that characterizes the nuclear suppression effect. The
implication on the parton energy loss in the context of recombination is
discussed.

\end{abstract}

\maketitle

The production of hadrons in Au-Au collisions at the
Relativistic Heavy-Ion Collider is found at high $p_T$
to depend sensitively on centrality \cite{de}.  At $\sqrt s=200$ GeV
the scaled inclusive cross section of pions at $p_T\approx$ 3-4 GeV/c
decreases by a factor of 4-5 when the centrality is varied from the most
peripheral to the most central \cite{ph,st}.  In a previous paper
\cite{hy} we reported the finding of a universal function \phii that
can describe all the inclusive cross sections at all
centralities. Such a scaling behavior is achieved by use of a
scaling variable $z$ that combines \pt with $N_{\rm part}$, the
number of participants.  In this paper we investigate the
origin of that scaling.  In particular, we consider the nature of
energy loss that can give rise to such a behavior.

Although the scaling behavior can be extended to include
energy dependence also \cite{hy2}, we restrict our
consideration here to centrality dependence only, and
emphasize the phenomenological implication of the data at
$\sqrt{s} = 200$ GeV, measured by PHENIX \cite{ph}.  We
shall not discuss energy loss at the parton level except near
the end, since perturbative QCD is not reliable for $p_T < 6$
GeV/c. Indeed, soft partons have been found to be important in
the hadronization process through recombination
\cite{hy3,fr,gr,li}.  We shall stay mainly at the hadronic
level that is phenomenological and uncontroversial, and
consider the nuclear suppression effect on the observed
hadrons.

Let us recall the scaling behavior found in \cite{hy}.  With
the definition of the variable
\begin{eqnarray}
z = p_T/K(N)
\label{1}
\end{eqnarray}
we find that the function $\Phi(z)$,
\begin{eqnarray}
\Phi(z)  = A(N) K^2(N) {1 \over 2\pi p_T} {dN_{\pi} \over
d\eta dp_T } \ ,
\label{2}
\end{eqnarray}
exhibits scaling behavior.  We have used the notation $N =
N_{\rm part}$, for brevity, and
\begin{eqnarray}
K(N) = 1.226 - 6.36 \times 10^{-4} N\ ,
\label{3}
\\
A(N) = 530 N_c(N)^{-0.9}, \qquad N_c(N)=0.44N^{1.33} \ ,
\label{4}
\end{eqnarray}
where both $K(N)$ and $A(N)$ are normalized to 1 at $N =
350$.  At all $p_T$ and $N$, the data of $dN_{\pi} / p_T
d\eta dp_T$ at midrapidity collapse to one universal curve $\Phi(z)$,
which is parametrized in \cite{hy} by
\begin{equation}
\Phi(z)=1200\ (z^2+2)^{-4.8}\ (1+25 e^{-4.5 z}) \ .  \label{4.5}
\end{equation}

It is clear that what gives rise to the scaling behavior must be related
to an universal property in the medium effect on the production of
pions.  Since it is not possible to determine experimentally  the
degradation of parton momentum as the medium size is increased, and
since whether hadronization is by means of fragmentation or
recombination is still controversial, we choose to stay at
the hadronic level and examine energy loss.  Since the produced pions
do not themselves traverse the dense medium, energy loss here does not
refer to the evolutionary process of a pion, as one can for a parton.
Instead, it refers to the shift of the pion distribution, as the medium
size quantified by $N$ is increased.

Let us now consider the implications of a scaling function
$\Phi(z)$.  Let $z$ be defined as
\begin{eqnarray}
z = x J(N) \ ,            \label{5}
\end{eqnarray}
where $x$ is a dimensionless momentum variable identified as
$x = p_T/p_0$ with $p_0$ chosen at $p_0 = 1$  GeV/c so that $x$ is
numerically the same as $p_T$.     We now examine the consequences of
writing \phii in terms of $x$ and $N$ explicitly
\begin{eqnarray}
\Phi(z) = F(x, N) \ .
\label{7}
\end{eqnarray}

In pQCD, such as in  Ref.\ \cite{ba}, one compares a distribution in
medium with one in vacuum in order to emphasize the medium
effect.  We prefer, however, to stay away from the $pp$ collision case,
since we want to
consider incremental changes of the medium size.  From our
perspective of dealing only with the observables, it is very natural to
ask the $\epsilon$-$\delta$ type question.  That is, given a medium that
is not too small, in which pions are produced, if its size is increased
by an $\epsilon$ amount, what is the corresponding downward shift
$\delta$ in momentum in order to maintain the same probability of
producing the pions?  In terms of $x$ and $N$, the proposition can
essentially be stated as
\begin{eqnarray}
F(x,N) = F ( x - \delta, N + \epsilon) \ .
\label{8}
\end{eqnarray}
The $+\epsilon$ and $- \delta $ relationship is a consequence
of the suppression effect.

For infinitesimal $\epsilon=\delta N$ and $\delta=\delta x$ we can
expand the right-hand side of Eq.\ (\ref{8}) and keep only the first
order terms, getting
\begin{eqnarray}
{\delta x \over \delta N}  = {\partial F / \partial N \over \partial
F / \partial x} \ .
\label{9}
\end{eqnarray}
If $F(x,N)$  satisfies Eqs.\ (\ref{5}) and (\ref{7}), then we
have
\begin{eqnarray}
{\delta x\over \delta N}  = {x \over J} {dJ \over dN} \ .
\label{10}
\end{eqnarray}
Since $\delta x$ is proportional to $x$, it is more sensible to consider
the fractional energy loss
\begin{eqnarray}
{\delta x \over x}  =  {d \ln J \over d \ln N} {\delta N \over
N} \ .
\label{11}
\end{eqnarray}
If the fractional energy loss is proportional to the fractional change of
centrality, a notion that seems extremely reasonable, i.e.,
\begin{eqnarray}
{\delta x\over x}  =  \gamma\ {\delta N \over N} \ ,
\label{12}
\end{eqnarray}
where $\gamma$ characterizes the suppression effect, then it is
necessary that
\begin{eqnarray}
J(N) = \left({N \over N_0
     } \right)^{\gamma}
\label{13}
\end{eqnarray}
for some normalization $N_0$.  Clearly, Eq.\ (\ref{12}) does not make
sense for $N$ very small, such as $N = 2$, since $\epsilon$ cannot be
made infinitesimal compared to 2.

The power-law behavior in Eq.\ (\ref{13}) is a necessary consequence of
scaling and fractional proportionality, Eq.\ (\ref{12}).
Comparing $J(N)$ with $1/K(N)$ in Eq.\ (\ref{3}), one finds that Eq.\
(\ref{13}) differs enough from our first scaling parametrization to cast
some doubt on whether Eq.\ (\ref{13}) is sufficient for scaling.
However, the fractional proportionality relationship is so compelling
that we have been motivated to reexamine the data, especially since the
original preliminary data have by now been finalized.  It should  be
noted that Eq.\ (\ref{13}) is obtained without relying on any specific
form for
$\Phi(z)$; it depends only on the structure of the scaling variable
expressed in Eq.\ (\ref{5}).  Thus one expects Eqs.\ (\ref{12}) and
(\ref{13}) to be very general properties of centrality scaling.

\begin{figure}[tbph]
\includegraphics[width=0.45\textwidth]{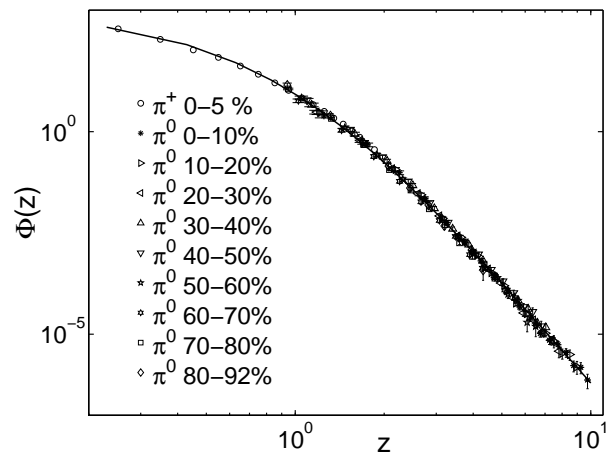}
\caption{ Scaling distribution $\Phi(z)$ showing the coalescence of
  all data points of 9 centrality bins for
$\pi^0$  and the most central collisions for $\pi^+$ production in Au+Au
collisions at $\sqrt s=200$ GeV as measured by PHENIX
\cite{ht,tc}. The solid line is a fit parametrized by Eq.\ (\ref{15}).}
\end{figure}

Using the data on $\pi^0$ from PHENIX tabulated on the web \cite{ht}
we have assembled the spectra for all measured centralities, and made
appropriate horizontal and vertical shifts in the log-log plot to
obtain a universal behavior.  The result is shown in Fig.\ 1. The
$\pi^+$ data for the most central collisions are used to supplement
$\pi^0$ in the low-$p_T$ region \cite{tc}.
The horizontal shift determines the scaling factor $J(N)$ shown in
Fig.\ 2(a). The vertical shift determines the normalization factor
$B(N)$ shown in Fig.\ 2(b).  The resultant scaling distribution
$\Phi(z)$ is now related to the measured inclusive distribution by
\begin{eqnarray}
\Phi(z) = {B(N) \over J^2(N)}{p_0^2 \over 2\pi p_T} {dN_{\pi} \over
d\eta dp_T} \ .
\label{14}
\end{eqnarray}
Evidently, the scaling behavior exhibited in Fig.\ 1 is very good.  The
solid line is a fit using the formula
\begin{equation}
\Phi(z) = 150\ (z^2 + 1.05)^{-4.18} \ (1 + 7e^{-4z} ) \ .
\label{15}
\end{equation}
There are many more points included in Fig.\ 1 than those of the
preliminary data used in Ref.\ \cite{hy}.

The behavior of $J(N)$ in the
log-log plot in Fig.\ 2 can be fitted by a straight line according
to Eq.\ (\ref{13}) with
\begin{equation}
\gamma = 0.077 \ .
\label{16}
\end{equation}
   The region $N<10$ is not considered. The normalization point is chosen
to be $N_0=325$, which corresponds to the most central 0-10\%
collisions \cite{ph}. The normalization factor $B(N)$ can be fitted by
two straight lines as shown by the solid lines. The
implication of this result will be discussed in the following.

\begin{figure}[tbph]
\includegraphics[width=0.45\textwidth]{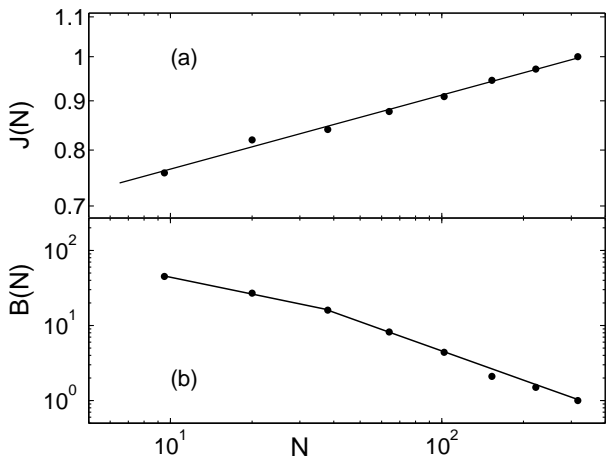}
\caption{(a) The scaling factor $J(N)$. Note that the vertical scale is
logarithmic. The solid line is a straightline fit, yielding
$\gamma=0.077$. (b) The normalization factor $B(N)$.
The solid lines are fits showing two scaling regions.}
\end{figure}

Although Eqs.\ (\ref{13}) and (\ref{16}) do not differ too much from
the inverse of $K(N)$ given in Eq.\ (\ref{3}), we have investigated the
source of the difference. Our conclusion is that in \cite{hy} we read
the preliminary data from a figure given in a conference talk \cite{dde}
and estimated the central points, whereas here we use the finalized data
in tabulated form, which differ slightly from the original. Thus our
present result is more reliable. Moreover, since in our new description
the medium suppression effect is characterized by one and only one
parameter $\gamma$, which plays the crucial role of specifying the
fractional proportionality relation in Eq.\ (\ref{12}), the resultant
scaling behavior has the distinction of being physically
motivated. As we have shown, it is not so much what the scaling function
\phii is as what the scaling variable $z$ is. Since $z$ quantifies the
difficulty of producing transverse motion, it can be termed {\it
transversality} that gives a universal description of that difficulty at
any centrality. The $z$ dependence of \phii in
Eq.\ (\ref{15}) differs little from that of Eq.\ (\ref{4.5}), except at
very high $z$ where the new data extend beyond that of the
preliminary result.

With the value of $\gamma$ in Eq.\ (\ref{16}) now determined
phenomenologically, we can return to Eq.\ (\ref{12}) and claim that
the notion of fractional proportionality has direct support by the
data.  Note that the independence of the fractional
energy loss $\delta x/x$ on $x$ is a result that differs from that of
pQCD  on the momentum shift of hard partons where
$\Delta p_T$ is proportional to $p_T^{1/2}$ \cite{ba}, and is
consistent with the independence of $R_{AA}$ on $p_T$ \cite{ph}.
Equation (\ref{12}) also implies that
$\delta x/\delta N$ is proportional to $x/N$ which is very different from
the assumption that the energy loss per unit length traversed by a
parton, $dE/dL$, is a constant. Although our phenomenological result on
the produced pions has no direct implication on the evolutionary
properties of the partons propagating through a medium, one should keep
these differences in mind, when the measurable consequences of what can
be calculated in pQCD are inferred.

Since the method by which we obtained $J(N)$ is by means of data
fitting, it is desirable to find an alternative method that is more
direct. Given $\Phi(z)$, one can determine the average
\begin{equation}
\left< z\right> = \int dz\ z^2 \Phi(z) \left/ \int dz\ z\,\Phi(z)\right.
= 0.42 \ .           \label{17}
\end{equation}
Then from Eq.\ (\ref{5}) we have $J(N)=0.42/\left< x\right>(N)$, where
$\left< x\right>$ is related to the inclusive cross section as
\begin{equation}
\left< x\right>(N)=\int dp_T\ p_T {dN_\pi\over d\eta dp_T}\left/ p_0\int
dp_T{dN_{\pi}\over d\eta dp_T} \right. \ .   \label{18}
\end{equation}
Since $J(N)$ is normalized to 1 at $N=N_0$, we now have
\begin{equation}
J(N)=\left< x\right>(N_0) / \left< x\right>(N)\ ,    \label{19}
\end{equation}
thus eliminating the reference to the scaling variable $z$.
   The $N$ dependence of $\left< x\right>(N)$ can be determined directly
from the data when the average is calculated at various centralities.
The large \pt part of the integration is important, for it is that part
of the $dN_{\pi}/d\eta dp_T$ that led us to the scaling factor $J(N)$ in
the first place. The low-\pt part of the spectra should be accurately
parametrized. Now, it is a matter of evaluating Eq.\ (\ref{18}) instead
of shifting and rescaling in Fig.\ 1. We recommend that experimental
groups that have the data, not only of $\pi^0$, but also of charged
hadrons, can use this method to check whether $J(N)$ indeed has the form
given in Eq.\ (\ref{13}).

The scaling behavior that we have obtained is for the produced
$\pi^0$. We have found in \cite{hy3} that the anomalously high $p/\pi$
ratio at $p_T\sim$ 3-4 GeV/c can be understood in terms of
recombination without fragmentation, when hard partons are allowed to
recombine with the  soft ones. That is possible if the recombination
function for the pion does not restrict the recombining
$q$ and $\bar q$ to have roughly the same momentum \cite{hhy}. Indeed,
in our view jet fragmentation is included in recombination because a
large-$p_T$ hard parton initiates a parton shower that hadronizes by
recombination \cite{rh}. The use of fragmentation function is only a
phenomenological way of parametrizing that process, and does not stand
for an independent hadronization mechanism. In heavy-ion collisions the
parton shower on the surface has low-$p_T$ components that mingle with
the soft partons with hydrodynamical origin. Since separating them would
be artificial, we treat them on equal footing in the recombination process
that can involve hard partons as well. The purpose of this discussion is
to prepare our way to descend to the parton level without fragmentation.
The scaling behavior that we have found supports this view, since our
rescaling procedure is universal and does not separate the high-\pt from
the low-\pt regions using different transversality variables. We note that
this view differs from that taken in \cite{fr}, in which recombination
and fragmentation are important in different regions. We also note that
our application of the recombination mechanism to all partons encounters
no inconsistency with the two experimental facts: (a) $p/\pi$ ratio is
roughly 1 at $p_T \sim$ 3-4 GeV/c, and (b) the jet structure in Au-Au
collisions is similar to that in $pp$ collisions in the same \pt region
\cite{jac}.

With the above discussion we have laid the basis for the expectation
that the quark distribution $F_q$ for all \pt contributes to the
determination of the pion distribution at all $p_T$. The
recombination equation derived in Refs.\ \cite{hy2,hy3} has the form
\begin{equation}
\Phi(z)=\int dz_1dz_2\,{z_1z_2\over z}F_q(z_1)F_{\bar
q}(z_2)\delta(z_1+z_2-z) \ ,    \label{20}
\end{equation}
in which the \pt variables have all been transformed to the scaling
variables with
\begin{equation}
z_i={p_{i_T}\over p_0}J(N),  \qquad i=1,2\ .   \label{21}
\end{equation}
   The $\delta$ function comes from the recombination
function and guarantees the conservation of momentum. Note that $z_1$
and $z_2$ are integrated over all values; in particular, they are not
restricted to the region $z/2$. Indeed, since the quark distributions
fall rapidly with $z_i$ it is necessary for one $z_i$ to be large and
the other $z_i$ to be small in order to give the highest contribution to
\phii at large $z$. The essential remark we want to make now is that in
our formalism for hadronization the $q$ and $\bar q$ distributions
must have their $p_{i_T}$ scaled by the same factor as shown in Eq.\
(\ref{21}). Since the $J(N)$ in Eq.\ (\ref{21}) is the same as that for
the pion, we conclude that the fractional energy loss for the quarks
(and antiquarks) satisfies the same proportionality relationship as in
Eq.\ (\ref{12}). It should immediately be emphasized that these
partons are at the end of their evolution (hydrodynamical and/or
branching in showers) just before recombination. As mentioned earlier
the energy loss discussed here does not refer to the radiative energy
loss of a parton traversing a medium \cite{wang}. Our point is that  the
fractional energy loss of the partons at hadronization satisfies the
same property as for the pions. Since  in this formalism of
hadronization we have been able to obtain the correct
$p/\pi$ ratio \cite{hy3}, it follows that the produced protons should
have the same property  in fractional energy loss also. This prediction
should be checked experimentally by studying the proton spectra and
seeing whether centrality scaling can be achieved with the same $J(N)$.
We expect, however, the proton mass effect to break the scaling at low
$z$.

Our final remark is a speculative one. The universality of the single
exponent $\gamma$ for all centralities (except the very peripheral
collisions) raises the question whether a hot and dense medium is any
different from a less dense medium in its effect on pion production at
high $p_T$. If not, the high \pt spectra would not be a fruitful place
to find the signature of plasma formation. A possible escape from that
conclusion is the observation that the result on $B(N)$ in Fig.\ 2(b)
can be fitted by two straight lines, which can be parametrized by
\begin{eqnarray}
B(N)&=&(N/N_1)^{-\beta_1},  \qquad N<38,   \nonumber \\
    &=&(N/N_2)^{-\beta_2},  \qquad N>38 \ ,
\label{22}
\end{eqnarray}
where $\beta_1=0.744$ and $\beta_2=1.292$ with $N_1=1610$ and
$N_2=325$. It  suggests that there are two regimes of $N$,
requiring different exponents $\beta_1$ and $\beta_2$ to achieve
scaling. If this break can be ascertained by more detailed analysis at
different energies, then it offers a way out of the strict universality in
which there is no hint of any essential diffference between hot and cold
media, leaving no room for any signal for the formation of quark-gluon
plasma.

To summarize, we have found a relationship between the centrality
scaling behavior of the observed pion spectra and the fractional energy
loss of the pions that is proportional to the fractional centrality
change. That proportionality is specified by an exponent $\gamma$,
which characterizes the medium suppression effect. In the recombination
model the same value of $\gamma$ is valid for the fractional energy
loss of the light quarks just before hadronization.  The existence of
the  universal scaling function of transversality suggests that
there is no essential difference in how the low- and high-$p_T$
hadronization processes should be treated. The possibility of a
single exponent $\gamma$ that can be directly extracted from the data
to summarize the nuclear suppression effect offers a very succinct
description of a complicated dynamical process.  Universal behavior
in centrality may be broken by the existence of two scaling regions in
the normalization factor, thus providing the possibility of a
threshold for a new distinctive regime.

We are grateful to Dr.\ D.\ d'Enterria for informing us of the PHENIX
data and for very useful comments. We also thank Drs.\ P.\ Jacobs, B.\
M\"uller and X.-N.\ Wang for helpful communication. This work was
supported, in part,  by the U.\ S.\ Department of Energy under Grant No.
DE-FG03-96ER40972.

\end{document}